\documentclass[10pt,a4paper]{article}
\usepackage{amsmath}
\usepackage{amssymb}
\usepackage{amsfonts}
\usepackage{amstext}
\usepackage{amsbsy}
\usepackage[mathscr]{eucal}
\usepackage{graphicx}
\newcommand{\beq}{\begin{equation}}
\newcommand{\eeq}{\end{equation}}
\newcommand{\beqa}{\begin{eqnarray}}
\newcommand{\eeqa}{\end{eqnarray}}
\newcommand{\ket}[1]{| #1 \rangle}

\makeindex
\title{\Large\textbf{Quantum entanglement and wedge product}}

\author{\textit{ Hoshang Heydari}\\
        \small\textit{Institute of Quantum
Science, Nihon University,}\\
\small\textit{1-8, Kanda-Surugadai, Chiyoda-ku, Tokyo 101- 8308,
Japan }}

\date{}
\begin{document}

\maketitle \thispagestyle{empty}

\maketitle
\begin{abstract}
In this paper, I will derive a measure of entanglement that
coincides with the generalized concurrence for a general pure
bi-and three-partite state based on wedge product. I will show
that a further generalization of this idea to a general pure
multipartite state with more than three subsystems will fail to
quantify entanglement, but it defines the set of separable state
for such composite state.
\end{abstract}

\section{Introduction}
Quantum entanglement is one of the most interesting properties of quantum mechanics.
It has become an essential resource for the quantum information (including quantum
communication and quantum computing) developed in recent years, with some potential
applications such as quantum cryptography \cite{Bennett84,Ekert91}and quantum
teleportation \cite{Bennett93}.  Quantification of a multipartite state entanglement
\cite{Lewen00,Dur99} is quite difficult and is directly linked to linear algebra, geometry,
and functional analysis. The definition of separability and entanglement of a multipartite state
was introduced in\cite{Vedral97}, following the definition for bipartite states, given by Werner
\cite{Werner89}. One of widely used measure of entanglement of a pair of qubits, is entanglement
 of formation and the concurrence, that gives an analytic formula for the entanglement of formation
 \cite{Bennett96,Wootters98,Wootters00}.
In resent years, there have been some proposals to generalize this measure to general bipartite
state \cite{Uhlmann00,Audenaert,Rungta01,Gerjuoy} and to multipartite state
 \cite{Albeverio,Bhaktavatsala,Akhtarshenas}. In this paper,
I will derive a measure of entanglement that coincides with
 the generalized concurrence for a general pure
bi-and three-partite state based on an algebraic and geometric
point of view, using wedge product, a useful tool from multi
linear algebra, which is mostly used in differential geometry and
topology in relation with differential forms. For a multipartite
state with more than three subsystems, this idea will not
completely quantify the entanglement of such a quantum state, but
it does define set of separable state for any general pure
multipartite state. To make this note self-contained, I will give
a short introduction to multi-linear algebra.

%%%%%%%%%%%%%%%%%%%%%%%%%%%%%%%%%%%%%%%%%%%%%%%%%%%%%%%%%%%%%%%%
\section{Quantum entanglement}
In this section we will define separable state and entangled state as well as give some example of entangled state. Let us denote a general pure composite quantum system
$\mathcal{Q}=\mathcal{Q}^{p}_{m}(N_{1},N_{2},\ldots,N_{m})
=\mathcal{Q}_{1}\mathcal{Q}_{2}\cdots\mathcal{Q}_{m}$ with $m$ subsystems, consisting of a state \begin{equation}\label{Mstate}
\ket{\Psi}=\sum^{N_{1}}_{i_{1}=1}\sum^{N_{2}}_{i_{2}=1}\cdots\sum^{N_{m}}_{i_{m}=1}
\alpha_{i_{1},i_{2},\ldots,i_{m}} \ket{i_{1},i_{2},\ldots,i_{m}}
\end{equation}
 defined on a Hilbert space
\begin{eqnarray}
\mathcal{H}_{\mathcal{Q}}&=&\mathcal{H}_{\mathcal{Q}_{1}}\otimes
\mathcal{H}_{\mathcal{Q}_{2}}\otimes\cdots\otimes\mathcal{H}_{\mathcal{Q}_{m}}\\\nonumber
&=&\mathbf{C}^{N_{1}}\otimes\mathbf{C}^{N_{2}}\otimes\cdots\otimes\mathbf{C}^{N_{m}},
\end{eqnarray}where the dimension of $j$th Hilbert space is given  by
$N_{j}=\dim(\mathcal{H}_{\mathcal{Q}_{j}})$. We are going to use
this notation through this paper, i.e., we denote a pure pair of
qubits by $\mathcal{Q}^{p}_{2}(2,2)$. Next,let
$\rho_{\mathcal{Q}}$ denote a density operator acting on the
$\mathcal{H}_{\mathcal{Q}}$. Then $\rho_{\mathcal{Q}}$ is said to
be separable, which we will denote by $\rho^{sep}_{\mathcal{Q}}$,
with respect to the Hilbert space decomposition, if it can be
written as
\begin{equation}\label{eq:sep}
\rho^{sep}_{\mathcal{Q}}=\sum^N_{k=1}p_k
\bigotimes^m_{j=1}\rho^k_{\mathcal{Q}_{j}},~\sum^N_{k=1}p_{k}=1
\end{equation}for some positive integer N, where $p_{k}$ are positive real numbers and $\rho^k_{j}$ denote a density operator on Hilbert space
$\mathcal{H}_{\mathcal{Q}_{j}}$. If $\rho^{p}_{\mathcal{Q}}$ represents a pure state, then the quantum system is separable if
$\rho^{p}_{\mathcal{Q}}$ can be written as
$\rho^{sep}_{\mathcal{Q}}=\bigotimes^m_{j=1}\rho_{\mathcal{Q}_{j}}$,
where $\rho_{\mathcal{Q}_{j}}$ is a density operator on
$\mathcal{H}_{\mathcal{Q}_{j}}$. If a state is not separable, then it is called an entangled state. Some of the most important entangled states are called the Bell states or $\mathrm{EPR}$ states.

%%%%%%%%%%%%%%%%%%%%%%%%%%%%%%%%%%%%%%%%%%%%%%%%%%%%%%%%%%%%%%%%%%%%%%%%%%%%%

%%%%%%%%%%%%%%%%%%%%%%%%%%%%%%%%%%%%%%%%%%%%%%%%%%%%%%%%%%%%%%%%%%%%%%%%%%%%%%%%%%%%%
%%%%%%%%%%%%%%%%%%%%%%%%%%%%%%%%%%%%%%%%%%%%%%%%%%%%%%%%%%%%%%%%%%%%%%%%%%%%%%%%%%%%%
\section{Multi linear algebra}

In this section we will establish a relation between Wedge product and concurrence of a general multipartite state. Let us consider the complex vector spaces $\mathrm{V}_{1},\mathrm{V}_{2},\ldots,
\mathrm{V}_{m}$ be vector spaces, where
$\dim(\mathrm{V}_{j})=N_{j},  ~ \forall j=1,2,\ldots, m $. Then we define a tensor of type $(m,n)$ on
$\mathrm{V}_{1},\mathrm{V}_{2},\ldots, \mathrm{V}_{m}$ as follows
\begin{eqnarray}
\mathcal{T}^{m}_{n}(\mathrm{V}_{1},\mathrm{V}_{2},\ldots,
\mathrm{V}_{m})&=&
  \mathcal{L}(\mathrm{V}_{1},\mathrm{V}_{2},\ldots,\mathrm{V}_{m};
  \mathrm{V}^{*}_{1},\mathrm{V}^{*}_{2},\ldots,\mathrm{V}^{*}_{n})\\\nonumber&=&
\mathrm{V}_{1}\otimes\mathrm{V}_{2}\otimes\cdots\otimes\mathrm{V}_{m}
\otimes\mathrm{V}^{*}_{1}\otimes\mathrm{V}^{*}_{2}\otimes\cdots\otimes\mathrm{V}^{*}_{n},
\end{eqnarray}
where $\mathrm{V}^{*}_{j}=\mathcal{L}(\mathrm{V}_{j};\mathbf{C})~
\forall j=1,2,\ldots, m $ is the space of linear applications
$\mathrm{V}_{j}\longrightarrow\mathbf{C}$ and is called dual of
$\mathrm{V}_{j}$. For any basis $e_{i}$ of $\mathrm{V}_{j}$ and
$e^{j}$ the dual basis of $e_{i}$ defined by
$e^{j}(e_{i})=\delta^{j}_{i}$, we have the following linear representation
\begin{eqnarray}
T=T^{i_{1},i_{2},\ldots,i_{m}}_{j_{1},j_{2},\ldots,j_{n}}
e_{i_{1}}\otimes e_{i_{2}}\otimes\cdots \otimes e_{i_{m}}\otimes
e^{j_{1}}\otimes e^{j_{2}}\otimes\cdots \otimes e^{j_{n}}.
\end{eqnarray}
I.e., we have
$\mathcal{T}^{1}_{0}(\mathrm{V}_{1},\mathrm{V}_{2},\ldots,
\mathrm{V}_{m})=\mathrm{V}_{1}$ and
$\mathcal{T}^{0}_{1}(\mathrm{V}_{1},\mathrm{V}_{2},\ldots,
\mathrm{V}_{m})=\mathrm{V}^{*}_{1}$.  Let $S_{m}$ be group of permutations $(1,2,\ldots, m)$. Then we call the tensor
$v_{1}\otimes v_{2}\otimes\cdots \otimes
v_{m}\in\mathcal{T}^{m}_{0}(\mathrm{V}_{1},\mathrm{V}_{2},\ldots,
\mathrm{V}_{m})$ symmetric if
\begin{eqnarray}
v_{1}\otimes v_{2}\otimes\cdots \otimes v_{m}=v_{\pi(1)}\otimes
v_{\pi(2)}\otimes\cdots \otimes v_{\pi(m)},
\end{eqnarray}
for all $\pi\in S_{m}$. The space of symmetric tensor is denoted by $\mathcal{S}^{m}_{0}(\mathrm{V}_{1},\mathrm{V}_{2},\ldots,
\mathrm{V}_{m})$. Moreover, we call the tensor $v_{1}\otimes
v_{2}\otimes\cdots \otimes
v_{m}\in\mathcal{T}^{m}_{0}(\mathrm{V}_{1},\mathrm{V}_{2},\ldots,
\mathrm{V}_{m})$ skew-symmetric if
\begin{eqnarray}
v_{1}\otimes v_{2}\otimes\cdots \otimes
v_{m}=\epsilon(\pi)v_{\pi(1)}\otimes v_{\pi(2)}\otimes\cdots
\otimes v_{\pi(m)},
\end{eqnarray}
for all $\pi\in S_{m}$, where $\epsilon(\pi)$ is the signature of permutation $\pi$. The space of symmetric tensor is denoted by
$\Lambda^{m}_{0}(\mathrm{V}_{1},\mathrm{V}_{2},\ldots,
\mathrm{V}_{m})$. Furthermore, we have the following mapping
\begin{equation}\label{Wedge}
\begin{array}{ccc}
 \mathrm{Alt}^{m}: \mathcal{T}^{m}_{0}(\mathrm{V}_{1},\mathrm{V}_{2},\ldots,
\mathrm{V}_{m})&\longrightarrow
&\Lambda^{m}_{0}(\mathrm{V}_{1},\mathrm{V}_{2},\ldots,
\mathrm{V}_{m}) \\
  v_{1}\otimes v_{2}\otimes\cdots \otimes
v_{m}&\longmapsto & v_{1}\wedge v_{2}\wedge\cdots \wedge v_{m}
 \\
\end{array},
\end{equation}where $v_{1}\wedge v_{2}\wedge\cdots \wedge
v_{m}=\frac{1}{m!}\sum_{\pi\in
S_{m}}\epsilon(\pi)v_{\pi(1)}\otimes v_{\pi(2)}\otimes\cdots
\otimes v_{\pi(m)}$. For example for $m=2$ we have
\begin{equation}
\begin{array}{ccc}
 \mathrm{Alt}^{2}: \mathcal{T}^{2}_{0}(\mathrm{V}_{1},\mathrm{V}_{2})&\longrightarrow
&\Lambda^{2}_{0}(\mathrm{V}_{1},\mathrm{V}_{2}) \\
  v_{1}\otimes v_{2}&\longmapsto & v_{1}\wedge v_{2}=v_{1}\otimes
v_{2}-v_{2}\otimes v_{1}
 \\
\end{array}.
\end{equation}

\section{Measure of entanglement for general bipartite state based on wedge product}
 From the wedge product, we can define a measure of entanglement for a
 general pure bipartite quantum system $\mathcal{Q}^{p}_{2}(N_{1},N_{2})$ as
\begin{equation}\label{bicon}
\mathcal{C}(\mathcal{Q}^{p}_{2}(N_{1},N_{2}))=\left(\mathcal{N}\sum^{N_{1}}_{1\mu<\nu}C_{\mu,\nu}\overline{C}_{\mu,\nu} \right)^{\frac{1}{2}}
\end{equation},where $C_{\mu,\nu}=v_{\mu}\wedge v_{\nu}$,
$v_{\mu}=(\alpha_{\mu,1},\alpha_{\mu,2},\ldots,\alpha_{\mu,N_{2}}),
~v_{\nu}=(\alpha_{\nu,1},\alpha_{\nu,2},\ldots,\alpha_{\nu,N_{2}})$
and $\overline{C}_{\mu,\nu}$ denote the complex conjugate of
$C_{\mu,\nu}$. That is, for a quantum system
$\mathcal{Q}^{p}_{2}(N_{1},N_{2})$, if we write the coefficients $\alpha_{i_{1},i_{2}}$, for all $1\leq i_{1}\leq N_{1}$ and $1\leq i_{2}\leq N_{2}$ in form of  a $N_{1}\times N_{2}$ matrix as below
\begin{equation}\label{ggg}
   \left( \begin{array}{cccc}
      \alpha_{1,1} & \alpha_{1,2} & \cdots & \alpha_{1,N_{2}} \\
      \alpha_{2,1} & \alpha_{2,2} & \cdots & \alpha_{2,N_{2}} \\
     \vdots & \vdots & \ddots & \vdots \\
      \alpha_{N_{1},1} & \alpha_{N_{1},2} & \cdots & \alpha_{N_{1},N_{2}} \\
    \end{array}\right),
\end{equation}then the vectors $v_{\mu}$ and $ v_{\nu}$ refer to different rows of this matrix. As an example let us look at the quantum system
$\mathcal{Q}^{p}_{2}(2,2)$ representing a pair of qubits. Then an expression for a measure of entanglement for such state using the above equation (\ref{bicon}) is given by
\begin{eqnarray}\label{pairq}
\mathcal{C}(\mathcal{Q}^{p}_{2}(2,2))&=&\left(\mathcal{N}C_{1,2}C^{*}_{1,2}
\right)^{\frac{1}{2}}
\\\nonumber&=&\left(2\mathcal{N}|\alpha_{1,1}\alpha_{2,2}-\alpha_{2,1}\alpha_{1,2}|^{2}
\right)^{\frac{1}{2}}
\\\nonumber&=&2|\alpha_{1,1}\alpha_{2,2}-\alpha_{2,1}\alpha_{1,2}|,
\end{eqnarray} where $C_{1,2}$ is given by
\begin{eqnarray}
C_{1,2}&=&v_{1}\wedge v_{2}
\\\nonumber&=&
(\alpha_{1,1},\alpha_{1,2})\otimes
(\alpha_{2,1},\alpha_{2,2})-(\alpha_{2,1},\alpha_{2,2})\otimes
(\alpha_{1,1},\alpha_{1,2})
\\\nonumber&=&
(0,\alpha_{1,1}\alpha_{2,2}-\alpha_{2,1}\alpha_{1,2}
,\alpha_{1,2}\alpha_{2,1}-\alpha_{2,2}\alpha_{1,1},0)
\end{eqnarray}
 for $\mathcal{N}=2$. The measure of entanglement for the general bipartite state defined in equation (\ref{bicon}) coincides with the generalized concurrence defined in \cite{Uhlmann00,Audenaert,Rungta01,Gerjuoy} and in particular equation (\ref{pairq}) that gives the concurrence of a pair of qubits, first time defined in \cite{Wootters98,Wootters00}.

\section{General multipartite state and wedge product}
Unfortunately, I couldn't directly generalize this result, using equation (\ref{Wedge}), to a general pure multipartite state.
I.e., for a general three-partite quantum system
  $\mathcal{Q}^{p}_{2}(N_{1},N_{2},N_{3})$ with the coefficients $\alpha_{i_{1},i_{2},i_{3}}$, for all $1\leq
i_{1}\leq N_{1}$, $1\leq i_{2}\leq N_{2}$ and $1\leq i_{3}\leq N_{3}$, we cannot identify the vectors $v_{\mu}$, $ v_{\nu}$ and $ v_{\tau}$. But generally, it could be possible for a quantum system with even number subsystems, i.e., $m=2,4,\ldots$.
 To be able to define a expression for degree of entanglement of a general pure multipartite state, I will construct a "artificial" wedge product between each subsystem in a composite system as in the case of the bipartite state. So, let us consider the quantum system
$\mathcal{Q}^{p}_{m}(N_{1},N_{2},\ldots,N_{m})$ and define
\begin{eqnarray}
\mathcal{C}_{k_{1}l_{1},k_{2}l_{2},\ldots,k_{m}l_{m}}&=&
\alpha_{k_{1},k_{2},\ldots,k_{m}}\alpha_{l_{1},l_{2},\ldots,l_{m}}.
\end{eqnarray}
Then, a wedge product between each pair of subsystem takes the following form
\begin{eqnarray}
\mathcal{C}_{k_{1}l_{1},k_{2}l_{2},\ldots,k_{j}\wedge
l_{j},\ldots,k_{m}l_{m}}&=&
\alpha_{k_{1},k_{2},\ldots,k_{j},\ldots,k_{m}}\alpha_{l_{1},l_{2},\ldots,l_{j},\ldots,l_{m}}\\\nonumber&&
-\alpha_{k_{1},k_{2},\ldots,l_{j},\ldots,k_{m}}\alpha_{l_{1},l_{2},\ldots,k_{j},\ldots,l_{m}}.
\end{eqnarray}
Now, we can define a measure of entanglement as follows
\begin{eqnarray}
\mathcal{E}(\mathcal{Q}^{p}_{m}(N_{1},\ldots,N_{m}))
&=&\nonumber\left(\mathcal{N}\sum_{\forall K,L}\sum_{\forall
j}\mathcal{C}_{k_{1}l_{1},\ldots,k_{j}\wedge
l_{j},\ldots,k_{m}l_{m}}
\overline{\mathcal{C}}_{k_{1}l_{1},\ldots,k_{j}\wedge
l_{j},\ldots,k_{m}l_{m}}\right)^{\frac{1}{2}}
\\
&=&(\mathcal{N}\sum_{\forall K,L}\sum_{\forall
j}|\alpha_{k_{1},k_{2},\ldots,k_{m}}\alpha_{l_{1},l_{2},\ldots,l_{m}}
\\\nonumber&&-
\alpha_{k_{1},k_{2},\ldots,k_{j-1},l_{j},k_{j+1},\ldots,k_{m}}\alpha_{l_{1},l_{2},
\ldots,l_{j-1},k_{j},l_{j+1},\ldots,l_{m}}|^{2})^{\frac{1}{2}},
\end{eqnarray} where $j=1,2,\ldots,m$ and multi-index
$K=(k_{1},k_{2},\ldots,k_{m}),L=(l_{1},l_{2},\ldots,l_{m})$.
%%%%%%%%%%%%%%%%%%%%%%%%%%%%%%%
As an example, let us look at the general pure three-partite quantum system $\mathcal{Q}^{p}_{2}(N_{1},N_{2},N_{3})$. The above measure of entanglement gives

%%%%%%%%%%%%%%%%%%%%%%%%%%%%
\begin{eqnarray}
\mathcal{E}(\mathcal{Q}^{p}_{3}(N_{1},N_{2},N_{3}))
&=&\nonumber\left(\mathcal{N}\sum_{k_{1},l_{1};k_{2},l_{2};k_{3},l_{3}}
\sum^{3}_{j=1}\left|\mathcal{C}_{k_{1}l_{1},k_{j}\wedge
l_{j},k_{3}l_{3}} \overline{\mathcal{C}}_{k_{1}l_{1},k_{j}\wedge
l_{j},k_{3}l_{3}}\right|^{2}\right)^{\frac{1}{2}}\\\nonumber
&=&(\mathcal{N}\sum_{k_{1},l_{1};k_{2},l_{2};k_{3},l_{3}}
(\left|\alpha_{k_{1},k_{2},k_{3}}\alpha_{l_{1},l_{2},l_{3}}-
\alpha_{k_{1},k_{2},l_{3}}\alpha_{l_{1},l_{2},k_{3}}\right|^{2}\\\nonumber&&
+\left|\alpha_{k_{1},k_{2},k_{3}}\alpha_{l_{1},l_{2},l_{3}}-
\alpha_{k_{1},l_{2},k_{3}}\alpha_{l_{1},k_{2},l_{3}}\right|^{2}\\&&+
\left|\alpha_{k_{1},k_{2},k_{3}}\alpha_{l_{1},l_{2},l_{3}}-
\alpha_{l_{1},k_{2},k_{3}}\alpha_{k_{1},l_{2},l_{3}}\right|^{2})^{\frac{1}{2}}.
\end{eqnarray}
This measure of entanglement coincides with the generalized concurrence \cite{Albeverio} and is equivalent, but not equal, to our entanglement tensor based on joint POVMs on phase space \cite{Hosh2}.
%%%%%%%%%%%%%%%%%%%%%%%%%%%%%%%%%
For a general multipartite state, that is, for $m\geq 4$ this measure $\mathcal{E}(\mathcal{Q}^{p}_{m}(N_{1},\ldots,N_{m}))$ is not invariant under local operations. To show it, let us consider the quantum system $\mathcal{Q}^{p}_{4}(2,2,2,2)$. In this case we can have seven types of separability between different subsystems as below: It may be possible to factor $\mathcal{Q}_{1}$,
$\mathcal{Q}_{2}$,
 $\mathcal{Q}_{3}$, or
 $\mathcal{Q}_{4}$ from the composite system. To check this we need to make four different permutation of indices and it is exactly what the measure $\mathcal{E}(\mathcal{Q}^{p}_{4}(2,2,2,2))$ does. But there are other types of separability in this four-qubits state, namely if it is possible to factor out
$\mathcal{Q}_{1}\mathcal{Q}_{2}$,
$\mathcal{Q}_{1}\mathcal{Q}_{3}$,
$\mathcal{Q}_{1}\mathcal{Q}_{4}$,
$\mathcal{Q}_{2}\mathcal{Q}_{3}$,
$\mathcal{Q}_{2}\mathcal{Q}_{4}$, or
$\mathcal{Q}_{3}\mathcal{Q}_{4}$. These six possible factorizations can be reduced to three checks of separability since if we test for separability of, i.e.,
$\mathcal{Q}_{1}\mathcal{Q}_{2}$, we have simultaneously tested
$\mathcal{Q}_{3}\mathcal{Q}_{4}$. For these types of separability we do need to perform more than one simultaneous permutation of indices. We have discussed these separabilities in relation with
Segre variety \cite{Hosh3}.
%%%%%%%%%%%%%%%%%%%%%%%%%%%%%
\section{Conclusion}
In this paper, I have derived a measure of entanglement that coincides with
concurrence of a general pure bipartite state based on mapping of a tensor
 product space to an alternating tensor product space defined by a wedge
 product. Moreover, I derived an expression for degree of entanglement for a
 multipartite state, which again coincides with generalize concurrence for a
 general pure three-partite state, but this measure fail to quantify entanglement
 for a composite system with more than three subsystems. However, this expression
 defines a set of separable states of a general multipartite state.

\begin{flushleft}
\textbf{Acknowledgments:} The author acknowledge useful
discussions with Gunnar Bj\"{o}rk. The author also would like to
thank Jan Bogdanski. This work was supported by the Wenner-Gren
Foundations.
\end{flushleft}

\end{document}